# Building the Signature of Set Theory Using the *MathSem* Program

Luxemburg Andrey

UMCA Technologies, Moscow, Russia

andr_lux@mail.ru

**Abstract.** Knowledge representation is a popular research field in IT. As mathematical knowledge is most formalized, its representation is important and interesting. Mathematical knowledge consists of various mathematical theories. In this paper we consider a deductive system that derives mathematical notions, axioms and theorems. All these notions, axioms and theorems can be considered a small mathematical theory. This theory will be represented as a semantic net.

We start with the signature <**Set**;∈> where **Set** is the support set, ∈–is the membership predicate. Using the ***MathSem*** program we build the signature <**Set;** ∈, ∩, ∪, ×,⊂ >, where ∩ – is set intersection, ∪– is set union, × -is the Cartesian product of sets, and ⊂– is the subset relation.

## 1 Introduction

The term "knowledge representation" usually means representations of knowledge aimed to enable automatic processing of the knowledge base on modern computers, in particular, representations that consist of explicit objects and assertions or statements about them. We are particularly interested in the following formalisms for knowledge representation:

1. First order predicate logic;
2. Deductive (production) systems. In such a system there is a set of initial objects, rules of inference to build new objects from initial ones or ones that are already build, and the whole of initial and constructed objects.
3. Semantic net. In the most general case a semantic net is an entity-relationship model, i.e., a graph, whose vertices correspond to objects (notions) of the theory and edges correspond to relations between them.

## 2 Description of the Project

In this paper we describe a part of the project and a part of the interactive computer application for automated building of mathematical theories. We define a formal language (close to first-order predicate logic), and a deductive (production) system that builds expressions in this language. There are rules for building new objects from initial (atomic) ones and the ones already built. Objects can be either statement (predicates), or definitions (these could be predicates or truth sets of predicates). These objects are put into vertices of a net (graph). There are several types of relations between objects: logical, syntactic, quantificational, set-theoretical, semantical. The membership predicate is taken as the atomic formula. Rules for building new objects include logical operations (conjunction, disjunction, negation, implication), adding a universal or existential quantifier, and one more rule: building the truth set of a predicate. One can consider symbols denoting predicates and sets, and also the predicates and sets themselves (when an interpretation or model is fixed). One more rule allows substitution of an individual variable or a term for a variable. The execution of the algorithm will result in closed formulae (without free variables), and a collection of such formulae can be considered as a set of axioms of some theory. The question is whether this theory has a model. Further, when we have built a new formula, we can simplify it using term-rewriting rules and logical laws (methods of automated reasoning). In order to prove theorems one can apply well-known methods of automated reasoning (resolution method, method of analytic tableaux, natural deduction, inverse method), as well as new methods based on the knowledge of «atomic» structure of the formula (statement) that we are trying to prove. For a new formula written in the formal language a human expert (mathematician) can translate it into «natural» language (Russian, English etc.), thus we obtain a glossary of basic notions of the system. More complicated formulae are translated into natural language using an algorithm and the glossary. There are several approaches of constructing mathematics: set-theoretical, constructive (algorithmic), topos theory etc. Mathematical logic is also part of metamathematics. Thus, one can consider various logics (intuitionistic, modal etc.) when constructing mathematical theories. The deductive system constructed here is based on classical first-order predicate logic. The initial object is the membership predicate, and the derivations result into mathematical notions and theorems. The computer program (algorithm) builds formulae from atomic ones (makes the semantic net of the derivation). In our systems we take formulae (strings) as vertices of the semantic net. Formulae can be predicates or mathematical objects. The principal relations between elements of the net are those that correspond to the method of obtaining an element (a vertice or an object) of the net from others. If a formula is obtained as the union of two others, then it is connected with those two formulae with the inheritance (parent — child) relation. Another relation is established between a mathematical object and the predicate from which this object was generated (as the truth set of this predicate), though we could have given it the same name. In the case of substitution of objects into predicates and making unions of objects the corresponding relations are established, and they correspond to links between objects (formulae). Relations in the net can be classified as semantical, syntactic, logical, set-theoretical and quantificational. A syntax relation is a subformula relation. Logical relations include disjunction, conjunction, negation, implication, equality. Set-theoretical relations define relations between objects as sets. Quantificational relations are classified by the quantifier used (universal or existential). We shall use an approach to knowledge representation that links together traditional notions of semantic nets in the procedural approach framework. One option is to formulate the semantics of nets in terms of classical logic. But the semantics of classical logics doesn't reflect data representation issues. E. g., there is no difference between a rule of inference that can be used, and the rule of inference that should be used. Another option is to formulate the semantics of nets by means of programs (algorithms). In this case the semantics include the notion of behavior w.r.t. some operations. This is so called procedural semantics. Any data representation should interact with the program at some level. In systems based on the notion of semantic net, where nets are considered as data structures, one usually builds an interpreter for modifying the structure and extracting answers to queries.

## 3 Deductive (Production) Systems

By a deductive (production) system we understand a triple

DS=<$O_i$, $R_j$, $O_k$ >, where $O_i$ is a set of initial objects, $R_j$ are rules for building new objects, and $O_k$ is the set of objects constructed, $O_i \subset O_k$

$$R_j\colon {O_k}^n \to O_k$$

Examples: <axioms, rules of inference, theorems>,

<a line segment; Cartesian product, sewing by the border; a square, a torus, a sphere, etc.>.

## 4 Definability and Expressibility

From predicate logic we shall use the notions of predicate, satisfiable, true and false formulae, and interpretation.

The question of definability was first mentioned in A. Padoa's works in connection with relations between Euclidean and non-Euclidean geometries. Later the notion of definability was introduced by A. Tarski. Craig's interpolation theorem and Beth's theorem played an important role in definability theory. Craig and Beth in their works demonstrated a close relation between the notions of definability and derivability. As a result, a series of important questions concerning definability were reduced to well-known problems of the theory of logical derivation.

Consider a given first-order signature and its interpretation on a set D. We are going to define the notion of definable (using formulae from the given signature in the given interpretation) k-ary predicate.

Choose k variables $x_1, x_2 \dots, x_k$ . Consider an arbitrary formula P, such that all its parameters belong to the list $x_1, x_2 \dots, x_k$. The truth value of this formula depends solely on truth values of variables $x_1, x_2 \ldots, x_k$. Thus, we obtain a mapping $D^k \rightarrow \{T, F\}$, i. e., a k-ary predicate on D. This predicate is called the predicate expressed by formula P. Predicates that can be expressed in this way are called expressible. A predicate P on set D is called expressible via predicates $P_1$, $P_2$,..., $P_k$ on D, if we can construct a statement that contains only $P_1$, $P_2$, ..., $P_k$ and is true if an only if P is true. Of course, besides these predicates we can use logical connectives and quantifiers.

Any k-ary predicate $P(x_1, x_2 \ldots, x_n, x_{n+1}, \ldots, x_k)$ (where $x_1, x_2 \ldots, x_k$ are free variables) defines a set M, which contains only such tuples $< x_1, x_2, \ldots, x_n >$ that $P(x_1, x_2 \ldots, x_n, x_{n+1}, \ldots, x_k)$ is true. This set is usually denoted by

$$M(x_{n+1}, \ldots, x_k) := \{ < x_1, x_2 \ldots, x_n > | \ P(x_1, x_2 \ldots, x_n, x_{n+1}, \ldots, x_k)\}$$

and is called "the set of all $< x_1, x_2 \ldots, x_n >$ such that $P(x_1, x_2 \ldots, x_n, x_{n+1}, \ldots, x_k)$". M is the truth set of the predicate.

# 5 Description of Deductive System of *MathSem*

**Notations**

$x_i$-variables for elements of sets; $A_j$ - variables for sets; $P_i$ denote predicates; $M_i$, $R_i$ -denote sets of mathematical objects; $\wedge$ (&), $\vee$, $\neg$, => - logical connectives; $\forall$-universal quantifier; $\exists$- existential quantifier; $\in$-denotes membership; $< x_1, x_2, \ldots, x_n >$ stands for a tuple; $\{ x_1, x_2\}$ is a set defined by explicitly listing all its elements.

Here we define objects derivable in the deductive system. They are either formulae that denote predicates, or formulae that describe mathematical objects (notions).

The symbol ":=" (a colon and the equality symbol) is considered as a metasymbol.

The main idea. We have the membership predicate. Consider predicates that can be defined via the membership predicate, and truth sets of these predicates.

The following formulae are considered atomic: $P_0(x_i, A_j) := x_i \in A_j$

Rules for building new formulae:

1. Negation: $P_i := \neg P_j$
2. Grouping with logical connectives: $P_k := P_i \vee P_j$, $P_k := P_i \& P_j$
3. Quantification: for any free variable *x* in predicate $P_i$ one can build new predicates $P_j := (\forall x) P_i$, $P_k := (\exists x) P_i$
4. Consider a finite number of variables: $x_1, x_2, \ldots, x_n$. One can construct a string $< x_1, x_2, \ldots, x_n >$ it is a triple of variables.
5. Building a mathematical object (notion). Consider a predicate $P(x_1, x_2 \ldots, x_k)$ (here $x_1, x_2, \ldots, x_n$ are free variables,). One can build the set
   $M(x_{n+1}, \ldots, x_k) := \{ < x_1, x_2 \ldots, x_n > | \ P(x_1, x_2 \ldots, x_n, x_{n+1}, \ldots, x_k)\}$
   It is the truth set of the predicate.
   $< x_1, x_2 \ldots, x_n > \in M(x_{n+1}, \ldots, x_k) \Leftrightarrow P(x_1, x_2 \ldots, x_n, x_{n+1}, \ldots, x_k)$
   or

$P_0(< x_1, x_2 \ldots, x_n >, M(x_{n+1}, \ldots, x_k)) \Leftrightarrow P(x_1, x_2 \ldots, x_n, x_{n+1}, \ldots, x_k)$

6. Substitution of variables. We can substitute variables in predicate P or object M. Since mathematical objects are actually sets, we can substitute them for variables into predicates.

Note that the interpretations plays an important role here. Different interpretations give different semantic values for predicates and different truth sets. To avoid logical paradoxes one can introduce a hierarchy of sets (B. Russel's simple type theory); another option is to choose rules of building new sets that do not allow possible paradoxes.

## 6 Software Description

We build semantic nets using the VUE (Visual Understanding Environment) program, written in an American university and freely distributed. The MathSem program itself is being written by Vitaly Tatarintsev. As for now, only syntax is implemented, the work on semantics is in progress. Only parent-son relations take place.

In this program, complicated formulae are built from atomic ones «manually». The formulae built can be saved in a Word file along with their descriptions. One can also upload formulae from a Word file. Below one can find an example of building circa 30 formulae. Notably, all the signature of set theory is built from formulae with length (number of atomic formulae) not greater than two.

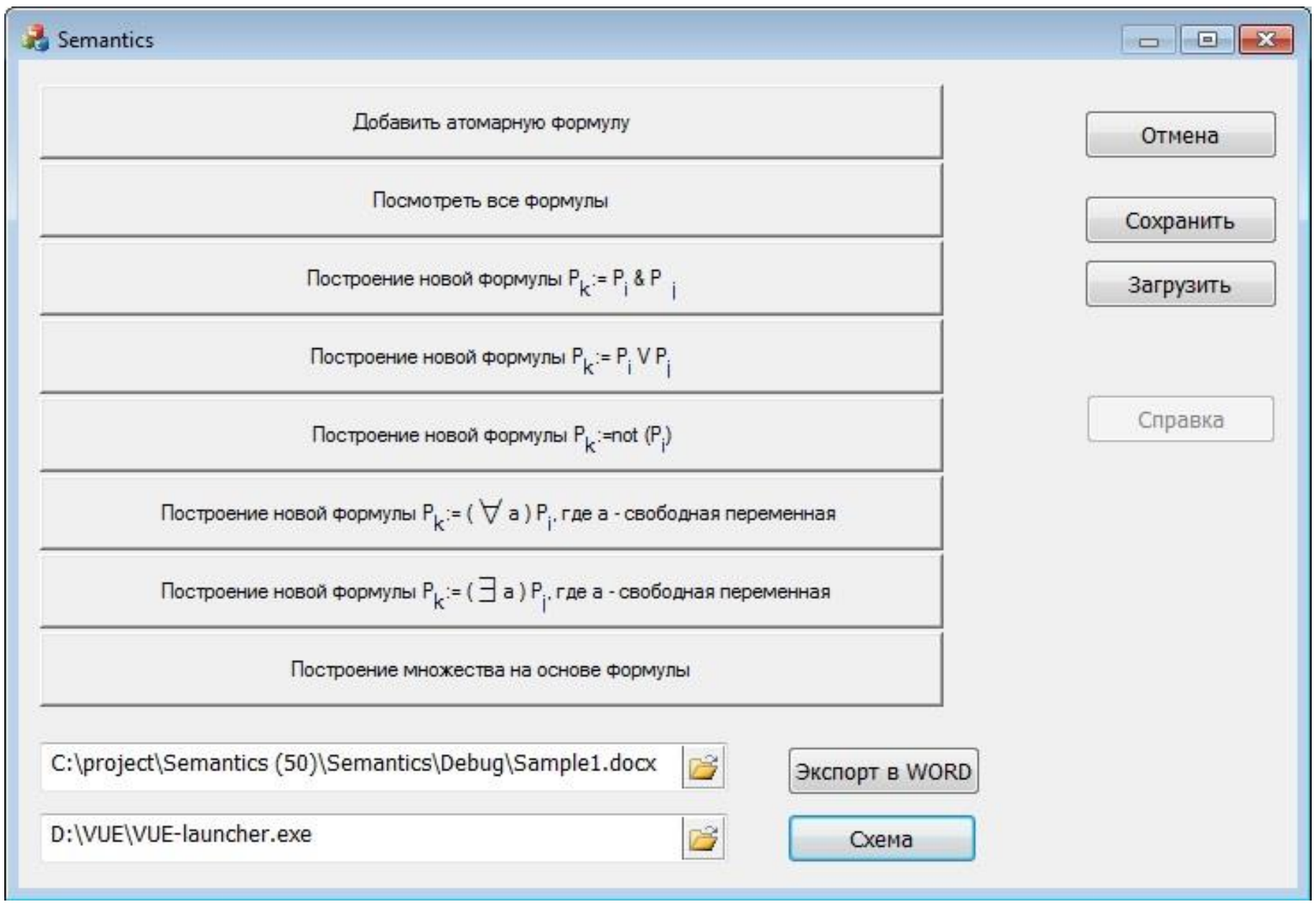

**Fig. 1.** First screen interface. Start dialog window.

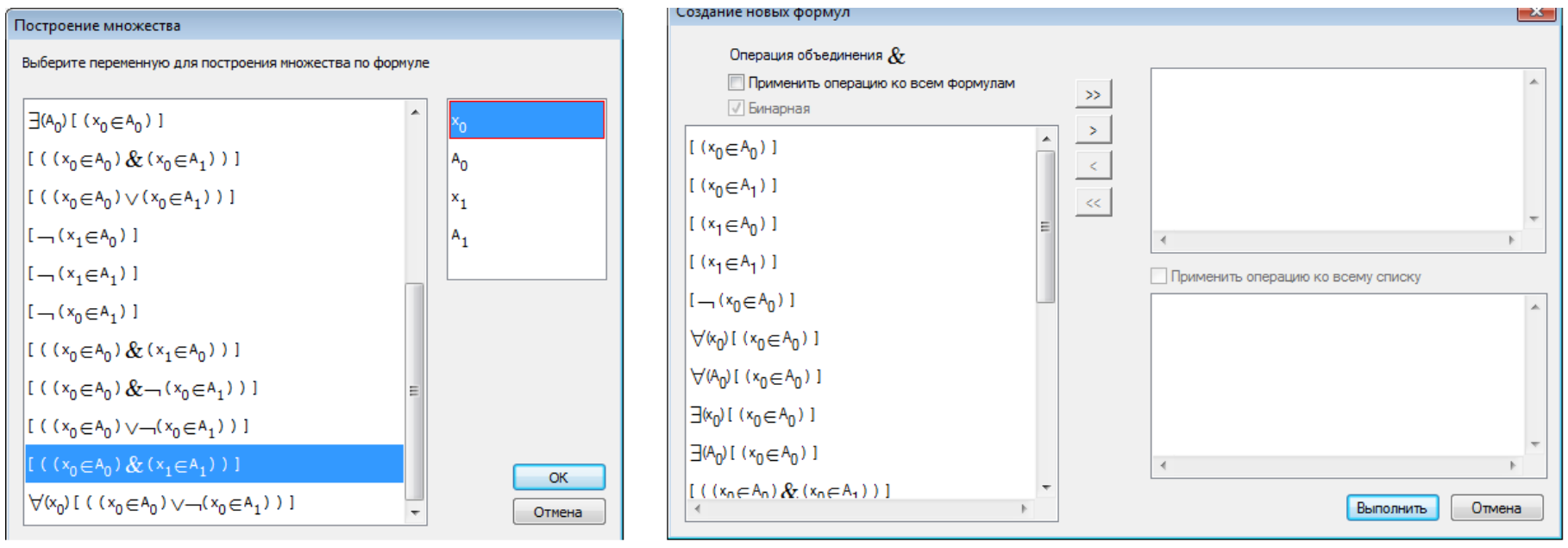

**Fig. 2.** Interfaces for building and viewing formulae

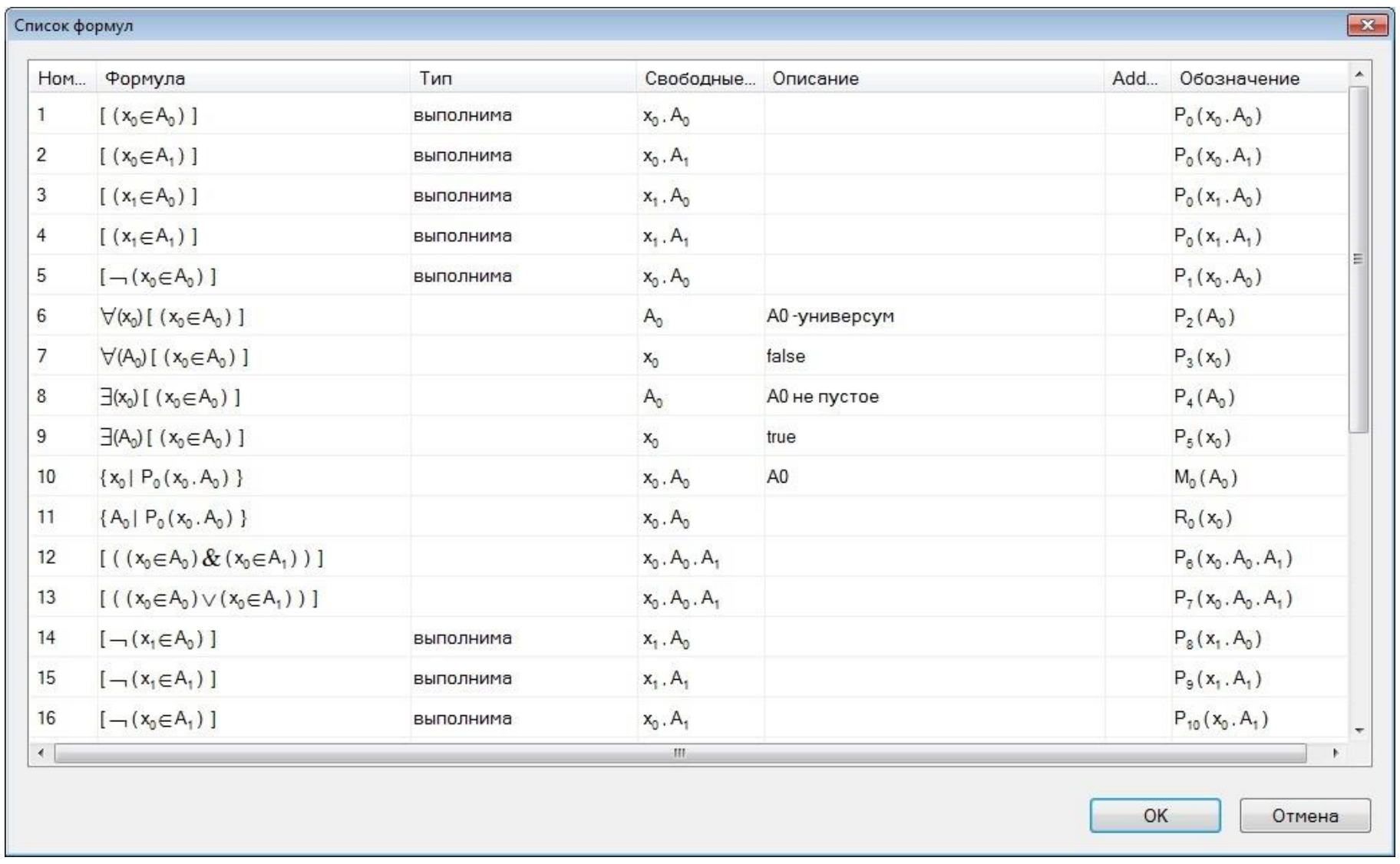

Список формул

| Ном... | Формула | Тип | Свободные... | Описание | Add... | Обозначение |
|---|---|---|---|---|---|---|
| 1 | $[\ (x_0 \in A_0)\ ]$ | выполнима | $x_0 . A_0$ | | | $P_0(x_0 . A_0)$ |
| 2 | $[\ (x_0 \in A_1)\ ]$ | выполнима | $x_0 . A_1$ | | | $P_0(x_0 . A_1)$ |
| 3 | $[\ (x_1 \in A_0)\ ]$ | выполнима | $x_1 . A_0$ | | | $P_0(x_1 . A_0)$ |
| 4 | $[\ (x_1 \in A_1)\ ]$ | выполнима | $x_1 . A_1$ | | | $P_0(x_1 . A_1)$ |
| 5 | $[\neg (x_0 \in A_0)\ ]$ | выполнима | $x_0 . A_0$ | | | $P_1(x_0 . A_0)$ |
| 6 | $\forall(x_0)[\ (x_0 \in A_0)\ ]$ | | $A_0$ | A0 -универсум | | $P_2(A_0)$ |
| 7 | $\forall(A_0)[\ (x_0 \in A_0)\ ]$ | | $x_0$ | false | | $P_3(x_0)$ |
| 8 | $\exists(x_0)[\ (x_0 \in A_0)\ ]$ | | $A_0$ | A0 не пустое | | $P_4(A_0)$ |
| 9 | $\exists(A_0)[\ (x_0 \in A_0)\ ]$ | | $x_0$ | true | | $P_5(x_0)$ |
| 10 | $\{x_0 \mid P_0(x_0 . A_0)\ \}$ | | $x_0 . A_0$ | A0 | | $M_0(A_0)$ |
| 11 | $\{A_0 \mid P_0(x_0 . A_0)\ \}$ | | $x_0 . A_0$ | | | $R_0(x_0)$ |
| 12 | $[\ (\ (x_0 \in A_0)\ \&\ (x_0 \in A_1)\ )\ ]$ | | $x_0 . A_0 . A_1$ | | | $P_6(x_0 . A_0 . A_1)$ |
| 13 | $[\ (\ (x_0 \in A_0) \vee (x_0 \in A_1)\ )\ ]$ | | $x_0 . A_0 . A_1$ | | | $P_7(x_0 . A_0 . A_1)$ |
| 14 | $[\neg (x_1 \in A_0)\ ]$ | выполнима | $x_1 . A_0$ | | | $P_8(x_1 . A_0)$ |
| 15 | $[\neg (x_1 \in A_1)\ ]$ | выполнима | $x_1 . A_1$ | | | $P_9(x_1 . A_1)$ |
| 16 | $[\neg (x_0 \in A_1)\ ]$ | выполнима | $x_0 . A_1$ | | | $P_{10}(x_0 . A_1)$ |

OK Отмена

**Fig. 3.** First part of table with formulae

Список формул

| Ном... | Формула | Тип | Свободные... | Описание | Add... | Обозначение |
|---|---|---|---|---|---|---|
| 14 | $[\neg(x_1 \in A_0)]$ | выполнима | $x_1 . A_0$ | | | $P_8(x_1 . A_0)$ |
| 15 | $[\neg(x_1 \in A_1)]$ | выполнима | $x_1 . A_1$ | | | $P_9(x_1 . A_1)$ |
| 16 | $[\neg(x_0 \in A_1)]$ | выполнима | $x_0 . A_1$ | | | $P_{10}(x_0 . A_1)$ |
| 17 | $[((x_0 \in A_0) \& (x_1 \in A_0))]$ | | $x_0 . A_0 . x_1$ | | | $P_{11}(x_0 . A_0 . x_1)$ |
| 18 | $[((x_0 \in A_0) \& \neg(x_0 \in A_1))]$ | | $x_0 . A_0 . A_1$ | | | $P_{12}(x_0 . A_0 . A_1)$ |
| 19 | $\{x_0 \mid P_{12}(x_0 . A_0 . A_1)\}$ | | $x_0 . A_0 . A_1$ | симметрическая разность A0 и A1 | | $M_1(A_0 . A_1)$ |
| 20 | $\{x_0 \mid P_6(x_0 . A_0 . A_1)\}$ | | $x_0 . A_0 . A_1$ | объединение A0 и A1 | | $M_2(A_0 . A_1)$ |
| 21 | $\{x_0 \mid P_1(x_0 . A_0)\}$ | | $x_0 . A_0$ | дополнение к A0 | | $M_3(A_0)$ |
| 22 | $\{x_0 \mid P_7(x_0 . A_0 . A_1)\}$ | | $x_0 . A_0 . A_1$ | пересечение A0 и A1 | | $M_4(A_0 . A_1)$ |
| 23 | $[((x_0 \in A_0) \vee \neg(x_0 \in A_1))]$ | | $x_0 . A_0 . A_1$ | | | $P_{13}(x_0 . A_0 . A_1)$ |
| 24 | $\{x_0 \mid P_{13}(x_0 . A_0 . A_1)\}$ | | $x_0 . A_0 . A_1$ | | | $M_5(A_0 . A_1)$ |
| 25 | $[((x_0 \in A_0) \& (x_1 \in A_1))]$ | | $x_0 . A_0 . x_1 . A_1$ | | | $P_{14}(x_0 . A_0 . x_1 . A_1)$ |
| 26 | $\{<x_0 . x_1> \mid P_{14}(x_0 . A_0 . x_1 . A_1)\}$ | | $x_0 . A_0 . x_1 . A_1$ | декартовое произведение A0 и A1 | | $M_6(A_0 . A_1)$ |
| 27 | $\forall(x_0)[((x_0 \in A_0) \vee \neg(x_0 \in A_1))]$ | | $A_0 . A_1$ | A1 подмножество A0 | | $P_{15}(A_0 . A_1)$ |
| 28 | $\{A_1 \mid P_{15}(A_0 . A_1)\}$ | | $A_0 . A_1$ | множество подмножеств A0 | | $R_1(A_0)$ |
| 29 | $\{<x_0 . x_1> \mid P_{11}(x_0 . A_0 . x_1)\}$ | | $x_0 . A_0 . x_1$ | | | $M_7(A_0)$ |

OK Отмена

**Fig. 4.** Second part of table with formulae

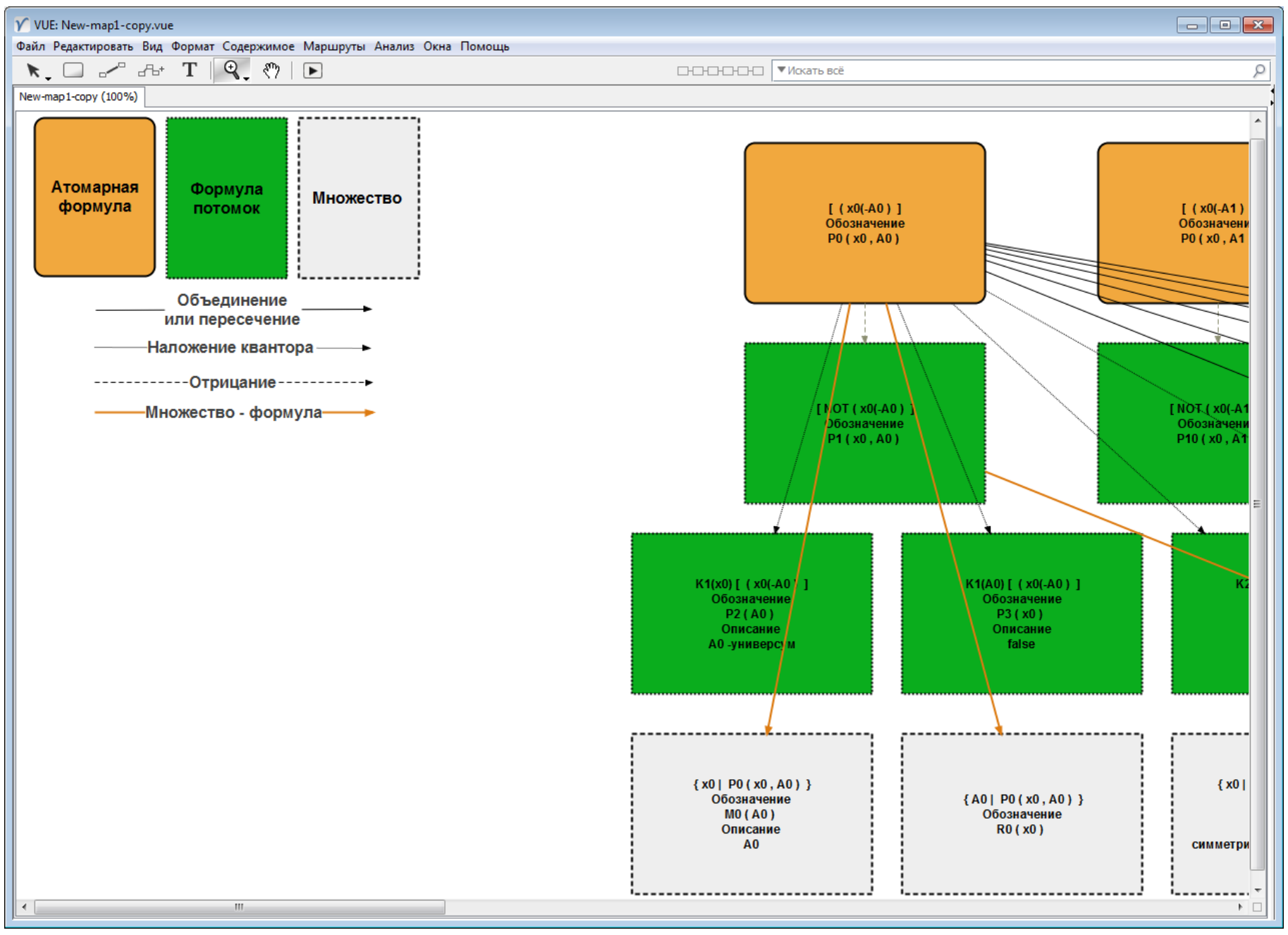

**Fig. 5.** Construction of the semantic net. In the first row there are atomic formulae. In the second row – their negations.

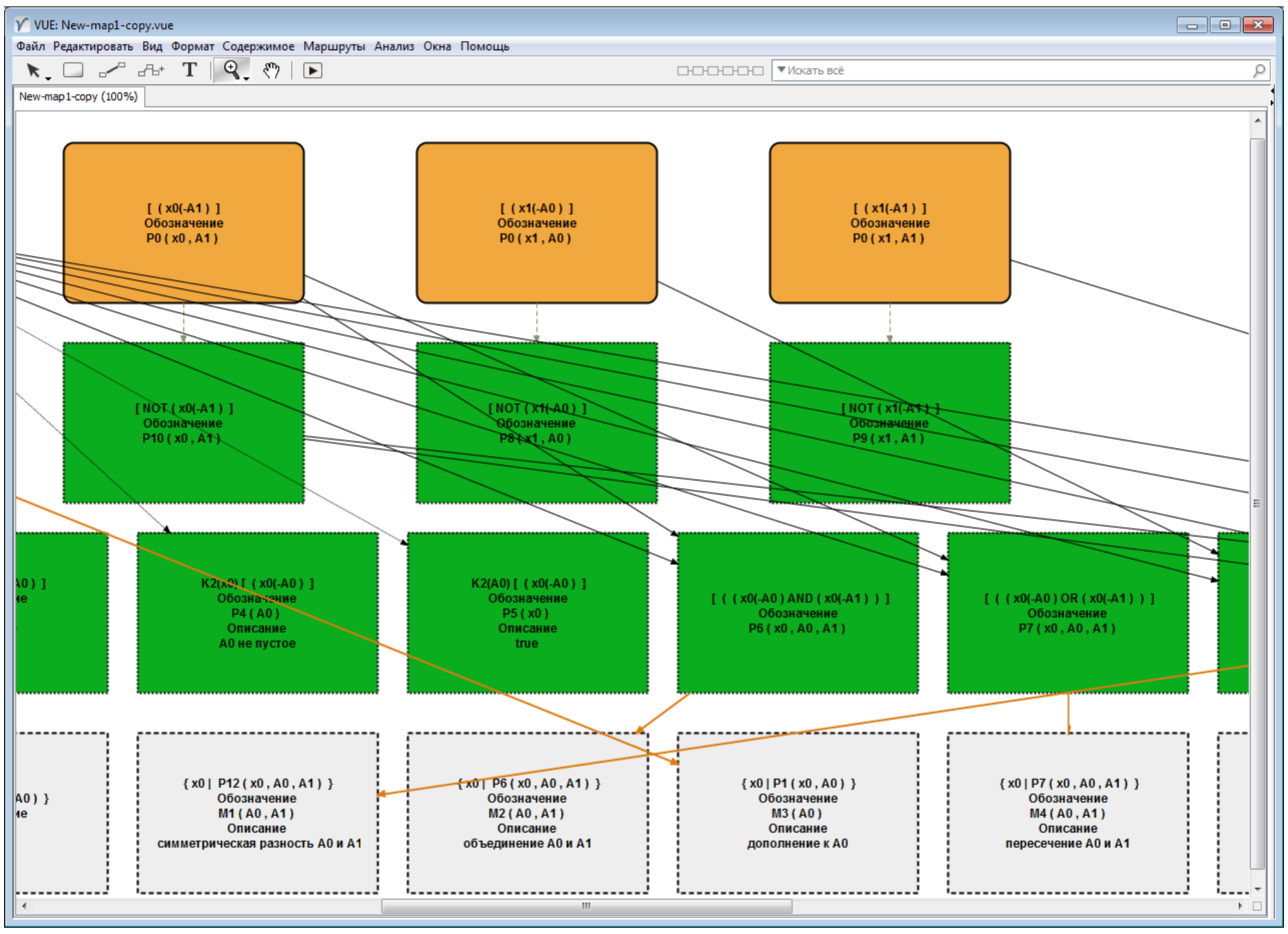

**Fig. 6.** Predicates and sets from elementary set theory

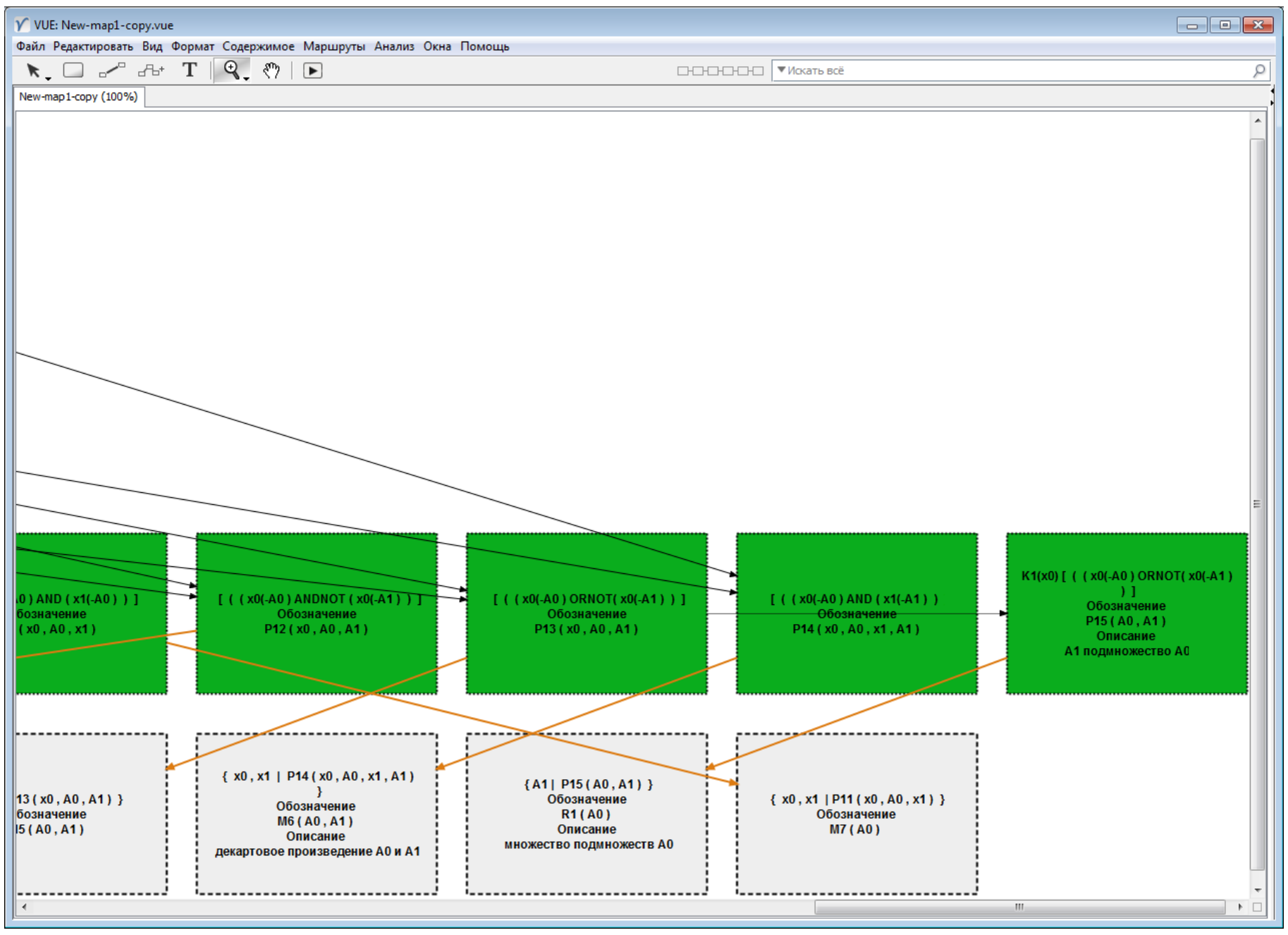

**Fig. 7.** Edges are parent-child relations

# 7 Table with result formulae.

| N | Формула / Formula | Обозначение/ Notation | Symbol | Natural language |
|---|---|---|---|---|
| 1 | $[\ (x_0 \in A_0)\ ]$ | $P_0(x_0, A_0)$ | | |
| 2 | $[\ (x_0 \in A_1)\ ]$ | $P_0(x_0, A_1)$ | | |
| 3 | $[\ (x_1 \in A_0)\ ]$ | $P_0(x_1, A_0)$ | | |
| 4 | $[\ (x_1 \in A_1)\ ]$ | $P_0(x_1, A_1)$ | | |
| 5 | $[\neg (x_0 \in A_0)\ ]$ | $P_1(x_0, A_0)$ | | |
| 6 | $\forall(x_0)\ [\ (x_0 \in A_0)\ ]$ | $P_2(A_0)$ | $A_0 = I$ | $A_0$ -universe |
| 7 | $\forall(A_0)\ [\ (x_0 \in A_0)\ ]$ | $P_3(x_0)$ | | |
| 8 | $\exists(x_0)\ [\ (x_0 \in A_0)\ ]$ | $P_4(A_0)$ | $A_0 \neq \varnothing$ | $A_0$ not empty set |
| 9 | $\exists(A_0)\ [\ (x_0 \in A_0)\ ]$ | $P_5(x_0)$ | | |
| 10 | $\{ x_0 \mid P_0(x_0, A_0) \}$ | $M_0(A_0)$ | $A_0$ | $A_0$ |
| 11 | $\{ A_0 \mid P_0(x_0, A_0) \}$ | $R_0(x_0)$ | | Ri are sets consisting of sets |
| 12 | $[\ ((x_0 \in A_0)\ \&\ (x_0 \in A_1))\ ]$ | $P_6(x_0, A_0, A_1)$ | | |
| 13 | $[\ ((x_0 \in A_0) \vee (x_0 \in A_1))\ ]$ | $P_7(x_0, A_0, A_1)$ | | |
| 14 | $[\neg (x_1 \in A_0)\ ]$ | $P_8(x_1, A_0)$ | | |
| 15 | $[\neg (x_1 \in A_1)\ ]$ | $P_9(x_1, A_1)$ | | |
| 16 | $[\neg (x_0 \in A_1)\ ]$ | $P_{10}(x_0, A_1)$ | | |
| 17 | $[\ (\ (x_0 \in A_0)\ \&\ (x_1 \in A_0)\ )\ ]$ | $P_{11}(x_0, A_0, x_1)$ | | |
| 18 | $[\ (\ (x_0 \in A_0)\ \& \neg (x_0 \in A_1)\ )\ ]$ | $P_{12}(x_0, A_0, A_1)$ | | |
| 19 | $\{ x_0 \mid P_{12}(x_0, A_0, A_1) \}$ | $M_1(A_0, A_1)$ | $A_0 \setminus A_1$ | difference of $A_0$ and $A_1$ |
| 20 | $\{ x_0 \mid P_6(x_0, A_0, A_1) \}$ | $M_2(A_0, A_1)$ | $A_0 \cap A_1$ | intersection of $A_0$ and $A_1$ |
| 21 | $\{ x_0 \mid P_1(x_0, A_0) \}$ | $M_3(A_0)$ | | the complement to $A_0$ |
| 22 | $\{ x_0 \mid P_7(x_0, A_0, A_1) \}$ | $M_4(A_0, A_1)$ | $A_0 \cup A_1$ | union of $A_0$ and $A_1$ |
| 23 | $[\ (\ (x_0 \in A_0) \vee \neg (x_0 \in A_1)\ )\ ]$ | $P_{13}(x_0, A_0, A_1)$ | | |
| 24 | $\{ x_0 \mid P_{13}(x_0, A_0, A_1) \}$ | $M_5(A_0, A_1)$ | | |
| 25 | $[\ (\ (x_0 \in A_0)\ \&\ (x_1 \in A_1)\ )\ ]$ | $P_{14}(x_0, A_0, x_1, A_1)$ | | |
| 26 | $\{ < x_0, x_1 > \mid P_{14}(x_0, A_0, x_1, A_1) \}$ | $M_6(A_0, A_1)$ | $A_0 \times A_1$ | Cartesian product of $A_0$ and $A_1$ |
| 27 | $\forall(x_0)\ [\ ((x_0 \in A_0) \vee \neg (x_0 \in A_1))\ ]$ | $P_{15}(A_0, A_1)$ | $A_1 \subset A_0$ | $A_1$ subset $A_0$ |
| 28 | $\{ A_1 \mid P_{15}(A_0, A_1) \}$ | $R_1(A_0)$ | | the powerset of $A_0$ |

**Table 1.** The $4^{th}$ column contains symbolic notation for extensions of the original language (signature).

Starting from the predicate of membership, we have obtained inclusion, intersection, union, complement, relative complement, Cartesian product of sets, subset relation, powerset.

<**Set**;∈>→ <**Set;** ∈, ∩, ∪, ×,⊂ >

We have built all the signature of set theory from the membership predicate in a combinatorial way. Usually in a first order language the signature is fixed in the beginning, so here we extend our language.

## 8 Connection between Predicates and Mathematical Objects

Using the fifth rule one can build a predicate from a set and a set from a predicate.

Given the definitions of union, intersection, and Cartesian product of sets, one can prove the following:

1. If $M_1=\{A_1|\ P_1(A_1,\ldots, A_k)\ \}$ and $M_2=\{B_1|\ P_2(B_1,\ldots, B_m)\}$, then $M_1xM_2=\{<A_1,B_1>|\ P_1(A_1,\ldots,A_k) \wedge P_2(B_1,\ldots, B_m)\}$
2. If $M_1=\{A_1|\ P_1(A_1,\ldots, A_k)\ \}$ and $M_2=\{A_1|\ P_2(A_1,\ldots,A_m)\ \}$, then $M_1\cap M_2=\{A_1|\ P_1(A_1,\ldots A_k) \wedge P_2(A_1,\ldots, A_m)\}$
3. If $M_1=\{A_1|\ P_1(A_1,\ldots, A_k)\ \}$ and $M_2=\{A_1|\ P_2(A_1,\ldots, A_m)\ \}$, then $M_1\cup M_2=\{A_1|\ P_1(A_1,\ldots, A_k) \vee P_2(A_1,\ldots, A_m)\}$.

We have built a semantic net of notions and statements of set theory. It is interesting to understand the connections between our system and ZFC. Some axioms of ZFC are syntactically derivable in our theory.